# How Theory-laden are Observations of Black Holes?

Juliusz Doboszewski[1]    Jamee Elder[2][1] †

[1]Black Hole Initiative - Harvard University, Cambridge, MA 02138, USA
[2]Department of Philosophy - Tufts University, Medford, MA 02155, USA

† Authors contributed equally.

February 14, 2025

## Abstract

We evaluate the roles general relativistic assumptions play in simulations used in recent observations of black holes including LIGO-Virgo and the Event Horizon Telescope. In both experiments simulations play an ampliative role, enabling the extraction of more information from the data than would be possible otherwise. This comes at a cost of theory-ladenness. We discuss the issue of inferential circularity, which arises in some applications; classify some of the epistemic strategies used to reduce the extent of theory-ladenness; and discuss ways in which these strategies are model independent.

## Contents

# 1   Introduction

In one important sense black holes are invisible as their gravitational pull is so strong that they can neither emit nor reflect any kind of radiation.[1] Nevertheless, several recent experiments have resulted in observations of black holes when they are coupled to either another compact object or matter accreting onto them (or, in further lines of evidence relying on microlensing, to a transient electromagnetic signal). The LIGO-Virgo Collaboration (LVC) "directly observed" black holes in 2015 through the detection of the gravitational waves produced when two black holes collide (B. P. Abbott et al. 2016b).[2] The Event Horizon Telescope (EHT) collaboration produced the first (also called "direct"[3]) images of (the "shadow" of) two supermassive black holes using a technique called Very Long Baseline Interferometry: M87* (The Event Horizon Telescope Collaboration et al. 2019a, 2019b, 2019c, 2019d, 2019e, 2019f, 2021a, 2021b) and Sgr A* (The Event Horizon Telescope Collaboration et al. 2022a, 2022b, 2022c, 2022d, 2022e,

---

[1]. With the possible exception of some quantum effects. For example, empirically unobserved emission of Hawking radiation is inversely proportional to the mass of a black hole. Assuming spherical symmetry, that would have to be below 0.00000002265 $M_\odot$ for its surface temperature to cross the cosmic microwave background threshold (at the current cosmological epoch approximately 2.725 K). Since the smallest known black hole candidates have masses above some $3M_\odot$, we will ignore emission of Hawking radiation in our discussion. Observation of other quantum effects, such as black hole superradiance (see Brito, Cardoso, and Pani (2020)), is more promising but has also not yet been made.

[2]. And many subsequent events. See R. Abbott et al. (2023) for the list of the most recent additions as well as pointers to catalogs of events from earlier observational runs.

[3]. Skulberg and Elder (2025) provide a critical analysis of the many senses of "directness" used in these detection claims.

2022f). Earlier experiments, such as the observation of stars orbiting the center of our own Milky Way galaxy, provide further empirical evidence of the existence of these objects ((Abuter et al. 2018)).[4]

In this paper we discuss how the LVC and EHT experiments gain empirical access to black holes, and the ways in which their conclusions utilize general relativistic simulations in an ampliative role. Ampliative use of simulations occurs when simulations allow one to extract more information about the phenomena from the observed data than would otherwise be possible (Jacquart 2020). The use of such simulations in drawing conclusions about black holes can introduce substantial theoretical assumptions into the methodology of these experiments, leading to concerns about the theory-ladenness of their result. In particularly troublesome situations, this may lead to inferential circularity, where conclusions drawn from observations depend on the very hypothesis being tested.

We begin by providing a primer on theory-ladenness and related concerns about circularity in section 2. Then in section 3 we discuss the role of numerical relativity simulations in developing the models used by the LVC to detect gravitational waves and observe black holes. We continue in section 4 by discussing the role of general relativistic magnetohydrodynamic simulations in drawing conclusions about the supermassive black hole candidate at the center of the Messier 87 galaxy (M87). Throughout our exposition we argue that both experiments feature simulations playing an ampliative role. In section 5 we then further argue that both of these experiments exhibit theory-ladenness as a result of the ampliative roles simulations play in them. We then assess the extent[5] to which each of these two observations should be considered theory-laden, and the ways in which this is (or is not) epistemically problematic. Some instances of this theory-ladenness lead to vicious circularity, but others do not. For example, the observation of black holes by the LVC (via parameter estimation for the sources of observed gravitational wave events) threatens with vicious circularity insofar as these observations are also used to test relativistic descriptions of the dynamics of these systems.

---

4. It should also be mentioned that defining or characterizing a black hole is a nuanced problem, as many inequivalent definitions of black holes are available. For example, a good evidence for the existence of black hole as defined by an apparent horizon might be insufficient to infer the existence of a black hole when it is defined by an event horizon. Due to complexity of the issue, here we take no stance on it, and suggest Curiel (2019) and Doboszewski and Lehmkuhl (2023) for a philosophical discussion and further references.

5. It is the most natural to talk about extent or degree of theory-ladenness, but quantifying it is extremely tricky; we will not attempt it.

In the context of historical sciences such as geology, paleontology, and geology, Currie (2018, 13) characterizes two positions about the prospects of these sciences: "[t]he pessimist predicts that our attempts to reconstruct the past will often fail; the optimist predicts that we will often succeed". Currie argues for the latter of these: optimism. By now, pessimism concerning astrophysics is a time-honored philosophical attitude.[6] Here is Hacking (1989, 577) expressing doubts concerning its methodology: "[t]he technology of astronomy and astrophysics has changed radically since ancient times, but its method remains exactly the same. Observe the heavenly bodies. Construct models of the (macro)cosmos. Try to bring observations and models into line. In contrast: the methods of the natural sciences have undergone a profound transformation, chiefly in the seventeenth century. Or one might say: the natural sciences came into being then and thereafter, while astronomy is not a natural science at all." The theory-ladenness of black hole observations could easily be interpreted in a pessimistic manner: if our only access to these highly unusual entities is highly theory-mediated, the prospects for obtaining solid evidence are bleak, because theory might bias us in their favor. We think there are good reasons not to endorse that attitude. Specifically, in section 6 we focus on a wide range of strategies used to mitigate epistemic risks associated with theory-ladenness and circularity. The availability of such strategies allows us to adopt an optimistic stance toward theory-laden observations of black holes, even in cases where theory-ladenness leads to inferential circularity. We articulate and classify these strategies and draw connections to notions of model independence—both in philosophical discussions of High Energy Physics and in astronomy.

Throughout our discussion, we draw on and elaborate Melissa Jacquart's account of the roles of simulations in astrophysics. In particular we extend Jacquart's discussion of amplifying role of simulations by analyzing cases where the ampliative role is *not* motivated by the problem of long timescales (as in Jacquart's example of collisional ring galaxies), but rather by other epistemic challenges of black hole astrophysics.[7]

---

6. In response, an emerging literature argues for a more optimistic attitude. For example, Anderl (2016), Jacquart (2020), and several contributions to Boyd et al. (2023) can be seen as defending optimism about the methods of astrophysics.

7. It is worth noting that Jacquart does not restrict the ampliative role to cases motivated by the problem of long timescales. However, she strongly emphasizes this problem as a key motivator for ampliative use of simulations and devotes most of her discussion to an example along these lines. In discussing alternative motivations in this paper, we take ourselves to be extending Jacquart's account of ampliative

# 2 Theory-ladenness: When is it a problem?

Philosophers have raised a range of worries about the ways in which observations are theory-laden. In this paper we are focusing on the theory-ladenness of measurement outcomes—the ways in which the "observations" derived from an experiment or measurement are affected by theoretical assumptions that are made in the course of the experiment. The greatest concern is that theory-ladenness may lead to a vicious circularity, where the theoretical assumptions made by the experimenters—either in the physical design of the experiment or in subsequent inferences from the empirical data—either make it more likely or even guarantee that the observation will confirm the theory being tested. This might occur if the hypothesis being tested is presupposed by the experimental methods being used. Franklin (2015, 156) provides the following toy example of this: if we use a mercury thermometer to test a hypothesis that substances expand as their temperature increases, this is theory-laden in a way that risks vicious circularity because the proper functioning of a mercury thermometer relies on the very hypothesis being tested.

One way that theoretical assumptions become embedded in experimental methodology is through computer simulations. Jacquart (2020) points out that simulation often becomes intertwined with observation in astrophysics due to challenging aspects of the epistemic situation of this science. In particular, the impossibility of controlled experiments on astrophysical targets (stars, galaxies, etc.) places constraints on how astronomers can investigate these systems. Simulations can help make up for the lack of controlled interventions by providing a proxy system that one can intervene on.

Jacquart (2020) describes three roles that simulations can play in the context of astrophysical observations: in hypothesis testing, exploring possibility space, and in amplifying observations. Simulations play all of these roles in the two cases we examine below.

However, for the purposes of this paper, we are most interested in the ampliative role of simulations. Jacquart describes this as follows:

> Amplifying observations occurs when the output of a simulation provides a
> new and sometimes unexpected context in which to interpret the data

---

simulation insofar as we are identifying and analyzing novel ways that these are important, and their relationship to other concepts, such as inferential circularity and model independence. But we take this to be compatible with Jacquart's original use of concept.

> present in the observation. [...] The simulations provide a means by which to learn about features of the target system that they did not know about before, thus enhancing the information derivable from the observations. (Jacquart 2020, 1217–8)

In other words, simulations play an ampliative role when they provide scientists with information that allows them to infer more from the data than they would otherwise be able to. This extends the scope of what the data can be taken to be informative about.

Jacquart provides an example of an ampliative use of simulations drawn from the case of collisional ring galaxies. In this case, simulations show that spokes are a short-lived feature of some of these systems. As a result, the spokes observed in some collisional ring galaxies can be used as a stand-in for system evolution over time. Since dynamics of these systems takes place over long timescales, collected data effectively provide only static snapshots of their evolution. These simulations, then, play an ampliative role: if an observed source displays spokes, additional information about timescale (in particular, time elapsed from the formation of the galaxy) can be extracted from the observational data. In this case, the ampliative use of simulations does not lead to inferential circularity, because the observation of spokes does not rely on the theoretical assumptions that go into the simulations. However, other instances of amplifying simulations can result in inferential circularities. In this paper, we take up the task of considering situations in which simulations are used ampliatively in black hole observations, and evaluating when the theory ladenness associated with that role leads to potentially-vicious circularity.

Before turning to our main case studies, it is worth spending a bit more time unpacking what we mean by an inferential circularity; when theory ladenness leads to such circularity; and the conditions under which such circularities are epistemically problematic.

In recent work, Ritson and Staley (2021) (following Beauchemin (2017)) consider the relationship between theory ladenness and what they call "evidential circularity". Under their characterization, "[t]he problem of evidential circularity arises when a measurement result is used to make an evidential argument in support of a theoretical claim that has been assumed in the production of the measurement result itself" (p. 155). (And, indeed, a very similar pattern will arise in our case studies of sections 3 and 4.) In a similar vein, we will characterize an inferential circularity as an instance of circular reasoning where a theoretical assumption T plays a constitutive role in a measurement (understood as a model-based inference) and where the resulting measurement outcome M is used as evidence in a subsequent inference to T (e.g., that T is valid). This formulation differs from Ritson and Staley (2021)'s (quoted above) in emphasizing the particular role that T plays in the measurement process, though we take this emphasis to be consistent with their overall discussion of circularity. In saying that T plays a constitutive role, we want to signal that close attention needs to be paid to the overall theory-testing procedure. The fact that the theory or hypothesis under test is implicated in an observation does not by itself imply inferential circularity. Instead, evaluating whether a given instance of theory-ladenness leads to inferential circularity depends on both the roles the theoretical assumptions play in generating empirical results and the wider experimental context.

We also follow recent work in the philosophy of measurement (especially Tal (2012, 2016)) which casts measurement as a model-based inference, emphasizing that a measurement outcome is a product of both a physical interaction between the measuring device and the target and subsequent inferences based on models of the measurement process. On this view, model- or theory-mediated measurement is the rule, rather than an exception. Theory ladenness, then, is a generic feature of measurement. With this in mind, the mere fact that a measurement is theory laden is insufficient reason to worry about the reliability of a measurement (at least not unless one endorses a general argument against the reliability of theory-mediated measurement, which we do not).

An example of benign theory ladenness is provided by Woodward (2011, 177–8)'s discussion of Millikan's oil drop experiment. Woodward argues that "[h]is experiment was such that he might well have obtained results showing that the charge of the electron was not quantized or that there was no single stable value for this quantity" despite contrary assumptions about the charge of electrons being made in the course of the experiment. Here, inferential circularity is avoided because the experiment could have provided evidence either for or against the theoretical assumption being made; the measurement outcome did not essentially depend on the truth of the theoretical assumption being tested. Smith and Seth (2020) make a similar point about Perrin's Brownian motion experiments, through which he measured the mean kinetic energy of the granules in the Brownian motion and (from these) obtained convergent values for Avogadro's number.

Theory-mediated measurement can even be a source of high quality evidence for the very theory implicated in the measurement. In the case of Newtonian celestial mechanics, Smith (2014) shows how an iterative process of accounting for anomalies between calculations and observations provided increasingly strong evidence for Newtonian mechanics over time. The basic procedure described by Smith is one where a small anomaly in an observed orbit is accounted for by incorporating the effects of further body into the calculation—a process he calls "closing the loop". Over time, observational precision increases and so the number of celestial bodies that (must) figure in the calculation increases. On this picture of theory-testing, the very theoretical assumptions that are constitutive of a research program accrue support by the success of that program in continuing to close the loop over time.

However, under different circumstances, theory-ladenness observations can be problematic. First, theory ladenness is a problem when the role of theoretical assumptions within the experiment either eliminates the possibility of outcomes that conflict with those assumptions, or else strongly biases the outcomes such that a disconfirming outcome is very unlikely (regardless of whether the assumption is true). This kind of bias occurs in instances of inferential circularity. Second, theory ladenness is a potential problem when there is no independent check on the experiment to establish the reliability of whatever process relies on the assumption. In such cases, any bias due to the theory-ladenness may be impossible to disentangle and eliminate. In other words, there are no external means of breaking the inferential circularity.

In some cases theory-ladenness does not lead to inferential circularity and may be virtuous rather than vicious (e.g., George Smith's "closing the loop" case for Newtonian solar system mechanics). In others (e.g., LVC gravitational-wave detection, EHT mass measurements), there is theory-ladenness with potential for inferential circularity that is rendered relatively benign by the availability of an independent check on the results of the theory-laden method. Here, we argue that theory-ladenness is potentially problematic, but under experimental control. The final case is one where there is inferential circularity without an independent check (e.g., LVC parameter estimation and theory testing). This is the most challenging scenario, where addressing the bias introduced by theory-ladenness is the most difficult. However, some mitigation strategies may still be available; we discuss these in section 6.

Let us briefly return to the thermometer case considered at the beginning of this section. In that case there was an inferential circularity when we considered the situation in isolation, because the hypothesis under test (mercury expands when heated) was crucial to inferring the measurement outcome (that the expansion of mercury corresponded to an increase in temperature). However, the circularity is easily resolved:

calibrating the mercury thermometer against a different kind of thermometer can establish the reliability of a mercury thermometer without pre-supposing the validity of the hypothesis under consideration.[8] In the remainder of this paper, we evaluate the extent to which inferential circularities emerge in observations of black holes, through the general relativistic assumptions that enter into them. We also consider the strategies available for mitigating these circularities in each case.

At this point, it is worth noting that although some of the observations of black holes we are discussing are used as tests of general relativity, they are at least equally often used to test other astrophysical hypotheses — some of which might rely on general relativistic assumptions. Forms of theory-ladenness which are problematic for the first purpose can be entirely innocuous for some of the latter purposes. Let us briefly consider two examples. Consider the situation of someone who is concerned whether current evidence is sufficiently strong to infer that Sgr A* is a supermassive black hole, and would like to utilize the framework of Table 3 of Eckart et al. (2017) (the table is a list of various necessary conditions for a source to be a supermassive black hole (SMBH), which are then further combined into various sufficient conditions). For that inferential purpose, the reliance of EHT Sgr A* mass estimates on general relativistic magnetohydrodynamic (GRMHD) simulations becomes problematic (as it would likely result in inferential circularity), although other ways of using EHT observations (for example, with future observations of photon rings) might not be as problematic in that regard. But for some other purposes certain general relativistic assumptions might be innocuous. For instance, black hole accretion is often modeled assuming a stationary background geometry, typically Kerr spacetime. Is that assumption problematic? As above: for tests of GR, yes, insofar as it might threaten them with inferential circularity. But for a more modest aim of investigating the process of accretion onto that particular source (and, for example, deciding whether it is a MAD or SANE type process)[9], not necessarily. The particular assumption about geometry of the source is widely shared in the field (indeed, Abramowicz and Fragile (2013) begin their overview of accretion with the view that "one of the main goals of the [black hole accretion disk] theory is to better understand the nature of black holes themselves", which can be understood as aiming to extract signatures of black holes through accretion), but it is not indispensable. GRMHD has been studied for objects other than black holes; it also remains falsifiable. For instance, it would fail if there was no central brightness depression in the EHT images; if that turned out to be the case, the field would likely enter a rapid period of setting up alternative GRMHD models.

---

8. Although this sounds trivial, Chang (2004) demonstrates that setting up a temperature scale alongside establishing reliable instruments for measuring this scale was a historically important and technically challenging undertaking.

9. Depending on the degree to which the accretion is magnetized, there are two modes in which accretion occurs: the SANE (standard and normal evolution) mode, where the gas pressure in the accretion dominates the midplane magnetic field pressure and the magnetic fields are turbulent; and the MAD (magnetically arrested disk) mode, where magnetic fields are strong, organized, and capable of disrupting accretion (The Event Horizon Telescope Collaboration et al. 2022e, 1). An important question to ask when learning about individual sources is whether the hot accretion flow of that source is MAD or SANE (or, possibly, something else).

With all this in hand, we are now ready to discuss our main case studies: observations of black holes with gravitational waves, and with radio interferometry.

# 3 LIGO-Virgo: Observation With Gravitational Waves

On September 14, 2015 the Laser Interferometer Gravitational-wave Observatory, (LIGO), comprising gravitational-wave interferometers in Hanford, WA, and Livingston, LA, detected gravitational waves for the first time.[10] The first detected signal, GW150914, is thought to have been produced by a binary black hole merger (B. P. Abbott et al. 2016b). This makes GW150914 the first "observation" of such a merging process, which encompasses a rapidly-decaying orbit of two black holes around one another (inspiral), followed by their collision (merger) and settling down to a stationary state (ringdown). Binary black hole mergers occur within the dynamical strong field regime of general relativity, where both high velocities and strong gravity come into play. The LVC observations provide the first (and for the foreseeable future only) empirical probes of this regime.

Gravitational-wave interferometers measure the strain of a passing gravitational wave through differential changes in the lengths of the two perpendicular interferometer arms.[11] These are highly sensitive instruments measuring a tiny effect; the change in the

---

10. At the time of GW150914 only the two advanced LIGO detectors had the sensitivity to make the detection. The current network, operated by the LIGO-Virgo-KAGRA Collaboration has expanded to include the Virgo detector in Italy and (more recently) the KAGRA detector in Japan.

11. Detailed information about all of these detectors can be found on their respective websites: https://www.ligo.caltech.edu/ (LIGO), https://www.virgo-gw.eu/ (Virgo), length produced by a passing wave is approximately $10^{-18}m$. These measurements are susceptible to many sources of noise. One of the main challenges of gravitational wave detection is recovering a gravitational-wave signal from data that is dominated by noise. The LIGO-Virgo Collaboration "observation" of GW150914 was a detection of gravitational waves using both modeled and unmodeled search pipelines (B. P. Abbott et al. 2016a).

The two modeled search pipelines, GstLAL and PyCBC, are targeted searches for gravitational waves produced by compact binary coalescence (e.g., by the merger of two black holes). Both searches use a signal-processing technique called matched filtering. This involves correlating a known signal, or *template* (derived using general relativistic simulations), with an unknown signal, in order to detect the presence of the template within the unknown signal. Using this technique to detect gravitational waves involves searching the data for templates corresponding to the full range of gravitational waves that might be present. The details of the modeled searches and their results for GW150914 are reported in B. P. Abbott et al. (2016a). Insofar as the observation of gravitational waves relies on matched filtering techniques, there is a clear sense in which this observation is theory- (or, perhaps more cautiously, model-) laden. The accuracy of these observations depends on having templates with morphologies that match real gravitational wave signals.

The unmodeled (in the sense that they do not use general relativistic templates) "burst" search algorithms, cWB and oLIB, look for transient gravitational wave signals by identifying coincident excess power in the time-frequency representation of the strain data from at least two detectors. Details of this search pipeline are reported in B. P. Abbott et al. (2016c). The burst search algorithms are not specialized for the detection of compact binary mergers and as such do not make strong assumptions about signal morphology. However, this lack of specialization means that they tend to report detections at lower statistical significance than modeled searches.

Agreement across the results of the modeled and unmodeled searches helps increase confidence in each. First, the use of the unmodeled search helps break the potential inferential circularity of the modeled search by demonstrating that the results do not rely on the theoretical assumptions built into matched filtering. Second, the modeled search helps validate results of the unmodeled search as genuine gravitational wave signals. The downside of building in fewer theoretical assumptions is that this makes the

---

https://gwcenter.icrr.u-tokyo.ac.jp/en/ (KAGRA).

burst search more susceptible to false positives caused by noise transients. In fact, one use of burst search algorithms is precisely to identify transients of loud noise to veto poor-quality data (Elder 2020, 18). This means that consistency with the results of the modeled search provides important evidence that a signal detected via the burst search represents a real gravitational wave.

Once a detection has been confirmed through the search pipelines, the next step is parameter estimation—an inference about the source system on the basis of the signal. The detected gravitational-wave signal can be thought of as a downstream trace of a (spatially and temporally) distant physical process that produced it.

Determining the properties of the source system is performed within a Bayesian framework. The basic idea is to calculate posterior probability distributions for the parameters describing the source system, based on some assumed model $M$ that maps parameters about the source system to gravitational-wave signals. The sources of the gravitational waves detected so far are compact binary mergers. Such events are characterized by a set of intrinsic parameters—including the masses and spins of the component objects—as well as extrinsic parameters characterizing the relationship between the detector and the source—including angular location in sky, luminosity distance, and orientation (relative to line of sight). The observation of compact binary mergers thus relies on having an accurate model relating the parameters describing the binary to the measured gravitational waveform. Inferring the parameters of the source from the gravitational wave signal GW150914 constituted a kind of "observation" of the binary black hole merger that produced it.

A range of modeling approaches are used for the modeled search pipelines and for parameter estimation. Of these, numerical relativity simulations are considered the most accurate source of information about the dynamics of the source system. These simulations take initial conditions of a binary black hole system and evolve them forward in time to extract the dynamical behavior of the system and its gravitational wave emission. These simulations act as benchmarks against which other modeling approaches are calibrated and tested.[12]

Both the detection of gravitational waves (via the modeled search pipelines) and the observation of black holes (via parameter estimation) by the LVC are examples of the use of simulations to amplify observations. As with the case of collisional ring galaxies, simulations provide new context for the interpretation of the data, allowing for drawing inferences about gravitational waves and their sources that would be impossible otherwise. Essentially, simulations unlock connections between the properties and dynamics of astrophysical sources (such as binary black hole mergers) and how these features are encoded in observational data. One way to see this amplifying role in action is to note that improved simulations, which account for additional physical effects (e.g., precession), lead to revised parameter estimates and thus the ability to "see" these additional effects in the data.

---

12. Elder (2023) offers a detailed analysis of the roles of various modeling approaches in the LVC observations. This includes an assessment of ways in which both the search and parameter estimation may be thought of as theory- or model-laden.

These cases are also examples of theory-laden observation, in the sense that the content of the observation depends on theoretical results (as derived from numerical relativity). Theoretical assumptions about the dynamics of binary black hole mergers are baked into the observation methods, and there is cause for concern about whether it is possible to make observations that are incompatible with these assumptions.

## 4  EHT: Observation with Very Long Baseline Interferometry

Active galactic nuclei (AGN) are highly luminous compact regions in galactic centers. Most non-stellar luminosity of AGN comes from accretion disks, which are expected to be hosted by supermassive black holes. An important milestone in empirically establishing the black hole paradigm for AGN is the image of the black hole in the center of M87 published in 2019 (The Event Horizon Telescope Collaboration et al. 2019a).

The EHT uses a process called Very Long Baseline Interferometry (VLBI), which involves using an array of stations separated by large distances (long "baselines") and the Earth's own rotation to create a high resolution virtual telescope. Pairs of components within the larger virtual telescope sample "fringe visibilities", which are the Fourier components of the source emission (Thompson, Moran, and Swenson 2017). Imaging with VLBI involves an inverse problem: modeling the source emission based on sampling of the visibilities. For M87*, three different methods of imaging the visibility data resulted in a thin ring with brightness asymmetry and a central dark region (The Event Horizon Telescope Collaboration et al. 2019d). The dark region was interpreted as containing the region of no escape, signaling the presence of a black hole.

Here we will not focus on the EHT's imaging process, in part because this process is not strongly theory-laden; for a detailed philosophical discussion of EHT imaging, see Doboszewski and Elder (2024). Instead, we look at the extraction of physical parameters of M87* via model-fitting. We further look at reconstruction of morphology and stability of M87* over a longer time period. We focus on M87* because the EHT has published extensively about that source and because its much lower characteristic timescales make it easier for the EHT to analyze than Sgr A* (which is the other main target of EHT's observations). However, many aspects of our discussion should generalize.

## 4.1 M87* black hole model fitting

How can one observe a region of no return? The black hole "shadow" is observable only if there are a sufficient number of photons emitted sufficiently near the black hole, and, moreover, if the surrounding plasma is transparent at the wavelength one observes. This is possible in an accretion process. Such emission processes are modeled through general relativistic magnetohydrodynamic simulations of magnetic field flow. Importantly, GRMHD simulations commonly assume that the gravitational source is described by the Kerr family of general relativistic spacetimes.

Physical black holes' parameters are extracted from the EHT data in three main ways (The Event Horizon Telescope Collaboration et al. 2019f):

1. Extracting image features (e.g., apparent ring size) and matching these to GRMHD models to extract best-fit physical parameters for the source.

2. Direct fitting of GRHMD simulations to visibility data.

3. Fitting geometric models to visibility data, then calibrating these models with GRMHD simulations. The models used include various types of Gaussians, rings, disks, and crescents. Of these, a generalized crescent family turns out to be preferable.

A crucial parameter of interest here is the angular size corresponding to one gravitational radius, $\theta_g$, which sets a physical scale for the source. The three methods converge to a common value of $\theta_g = 3.8 \pm 0.4 \mu$as, and consequently a source mass of $M = 6.5 \pm 10^9 M_\odot$. However, the EHT stress that "[a]ll of the individual $\theta_g$ estimates use the GRMHD simulation library [...] [a] degree of caution is therefore warranted. The measurements rely on images generated from GRMHD simulations and should be understood within that context" (The Event Horizon Telescope Collaboration et al. 2019f, 21).

The upshot of this is that GRMHD simulations, involving substantial general relativistic assumptions about the source's geometry, are shared across all three ways of estimating physical parameters in the 2019 EHT results. This makes the inferences about the source's parameters highly theory-laden.

## 4.2 Monitoring the dynamics of M87*

Proto-EHT arrays (2009, 2011, 2012, 2013) comprising 3 to 5 telescopes did not have sufficiently many baselines to resolve the source and produce its image. Even the 2013 data are equally compatible (by both Bayesian and Akaike information criteria) with a ring model of the previous section, and an asymmetric Gaussian model (which does not contain a central dark region). However, once the image resulting from 2017 data is used as a prior, the EHT can to some extent constrain evolution of the source's geometry, provide a partial answer about its stability, and as a result provide some evidence that the ring in 2017 observations is not a temporary feature of the source.

Wielgus et al. (2020) investigate whether the properties of geometric models can be constrained using a small dataset, such as the proto-EHT data. To do so, they begin by generating synthetic proto-array data—data that model what a smaller pre-2017 array would have observed, given a variety of source models. These source models include snapshots of GRMHD simulations, and models based on 2017 data.[13] Given a known source model, Wielgus et al. (2020) can test the accuracy of their methods before turning attention to pre-2017 datasets.

The method of Wielgus et al. (2020) involves fitting geometric models to each dataset to evaluate the values of various parameters (e.g., source diameter) describing the source. In particular, they focus on two models from the generalized crescent family: a concentric "slashed" ring and a thin slashed ring blurred with a Gaussian kernel. These two models have overlapping but nonidentical parameters; both measure the source diameter (d) and the orientation of any brightness asymmetry ($\phi_B$), but measure the presence of a central flux depression differently—see Wielgus et al. (2020, eq.5) for details. Applying this procedure to the 2009-2013 data, it turned out that while the 2009-2012 data had some preference for the central dark region, "only the 2013 archival data set provides a robust detection" of it (Wielgus et al. 2020, 17).

---

13. The GRMHD snapshot is processed through relativistic ray tracing, so that the photon paths are easier to see.

Although each of the pre-2017 datasets is consistent with a Gaussian model, Wielgus et al. (2020) point out that these "Gaussian models are very inconsistent in size, shape, and orientation across different years", whereas the best-fit ring models exhibit a high level of consistency across datasets. This further supports the hypothesis that the source was also a ring in 2009-2013 (as the alternative explanation becomes physically less plausible).[14] To be a bit more fine-grained, this modeling procedure provides evidence about which features of the source are preferred (a ring), consistent over time (ring diameter) and those that change over time (the position of maximum brightness).

In this case the initial analysis of synthetic data identifies stable features of a source, enabling the search for their signatures in the past observations. This enables the extraction of more information from the same data, which in turn places constraints on how the source changed over time. In other words, this is another instance of the amplifying role of astrophysical simulations in a dynamical regime. This case of the ampliative use of simulations relies on empirical input in the form of the 2019 image, and so functions in a feedback loop involving other observations.

In this case, the EHT Collaboration is able to extract more information from historical data by imposing stronger theoretical assumptions. By embracing increased theory-ladenness in the observation method, the EHT scientists can derive new conclusions—in this case, about the behavior of the source over time.

# 5 Discussion: The Pros and Cons of Ampliative Simulations

In Jacquart (2020)'s discussion the major motivation provided for the use of ampliative simulations, was a particular challenge of astrophysical observations: the fact that typical astrophysical targets evolve over very long timescales. This is certainly the case in her example, which concerns the use of simulations to extract dynamical information about the formation of spokes in ring galaxies, despite only being able to observe snapshots of such galaxies. However, as Jacquart acknowledges (but doesn't discuss in detail), the ampliative use of simulations can have other motivations.

---

14. 2018 data collected with the full EHT array also support the hypothesis that the ring persists, see The Event Horizon Telescope Collaboration et al. (2024).

In black hole astrophysics simulations are used ampliatively even in contexts where characteristic timescales of signals of interest are short. For LVC, the relevant timescales range from milliseconds (for large binary black hole mergers) to minutes (for binary neutron star mergers). For the main EHT sources the timescales range from minutes to months, with the most important timescale dictated by the period of the innermost stable circular orbit (ISCO): the time taken for light to travel around the black hole. For Sgr A* the period of the ISCO is 4-30 minutes (The Event Horizon Telescope Collaboration et al. 2022a). It also displays daily flares (Tiede et al. 2020). For M87* the period of the ISCO is 2.4–57.7 days (table 1 of The Event Horizon Telescope Collaboration et al. (2019b)). Secondary sources, such as 3C279 jet, also display day-to-day variability (Kim et al. 2020). (See also a more philosophical discussion of these issues in section 4 of Doboszewski and Lehmkuhl (2023).) With these characteristic timescales, the need for simulations to play an ampliative role comes down to other challenges that characterize the epistemic situation of black hole astrophysics. This includes theoretical challenges, such as the lack of any exact analytic solutions describing binary black hole mergers. It also involves more empirical challenges; for example, performing parameter estimation about a supermassive black hole is hampered by the fact that a radio telescope measures brightness of emission from some region, where the hot plasma is located. Even under the assumption that this region is a black hole, the region accessed by the EHT observations (which, in GR, corresponds to the "shadow", which in spherical symmetry is located at distance R=3M, with the event horizon at R=2M) is placed at a significant distance from the black hole's horizon. Therefore, GRMHD simulations connecting the underlying spacetime geometry and parameters of the black hole to the emission of matter in the accretion disc are needed.

Nevertheless, both the LVC and the EHT use simulations in an ampliative role in Jacquart's sense; simulations provide a new means of learning about features of the target system. In each case, simulations increase the scope of what it is possible to infer about the astrophysical black hole systems on the basis of the recorded data. But the use of simulations to amplify observations comes with the price of theory-ladenness. While this price need not always be costly, it warrants some caution. If the theory proves to be an inaccurate representation of the target system then theory-ladenness can lead to what Yunes and Pretorius (2009) call "fundamental theoretical bias", and be a source of systematic errors. In cases where that same theory is under test, theory-ladenness can even lead to inferential circularity.

Whether or not theory-ladenness leads to problematic biases or circularities depends on the details of the role that theoretical assumptions are playing in the observation, and the particular hypothesis being investigated. In the cases of complex experiments like LVC and the EHT, it can be challenging to discern when and for what purpose a particular instance of theory-ladenness is problematic. In the remainder of this section, we evaluate the extent to which each instance of theory-ladenness considered in this paper is epistemically problematic, given the particular inferential role that theory (via amplifying simulations) is playing in each case.

The LVC observation of gravitational waves via matched filtering is highly theory-laden, but this theory-ladenness looks to be relatively benign. In part, this is because the alternative, unmodeled pipeline provides an independent check on the results. The "burst" search successfully (though less confidently than the PyCBC and GstLAL pipelines) detected GW150914 without making the same theoretical assumptions about the dynamics of binary black hole mergers or the morphology of the corresponding gravitational waves. This shows that these assumptions are not necessary for the detection of gravitational waves. Indeed, the unmodeled searches not only provide an independent check, but also sometimes perform as well as or better than modeled searches; some examples include events GW191230_180458 and GW200225_060421, for which the calculated signal-to-noise ratio based on the unmodeled search was greater than for the modeled searches (R. Abbott et al. 2023). An additional consideration is the use of residuals tests to check whether the residual data (after the detected signal is removed) is consistent with Gaussian noise. This test places constraints on how much signal is missed by the model-based search methods (B. P. Abbott et al. (2016d); see also Elder (2023) for philosophical discussion of this test.)

LVC black hole observation (via parameter estimation) is also an instance of theory-ladenness: what can be observed is constrained by the possibilities determined by numerical relativity simulation. The properties attributed to the source system depend on the model $M$ used in mapping between the source parameters and a gravitational wave signal. This is an inherently model-dependent inference for which there is no agnostic, unmodeled alternative. If the numerical relativity results are inaccurate, either with respect to the underlying theory of general relativity or with respect to the actual behavior of such physical systems, then parameter estimation is correspondingly biased. Additionally, this theory-ladenness threatens to become an inferential circularity insofar as the LVC's observations are intended to confirm the predictions of general relativity for such systems. (See Elder (2023) for a detailed discussion of this circularity problem.) This circularity may be viewed as especially problematic given that no alternative empirical access to the target system seems to be possible. So (unlike in the thermometer case above) the circularity is difficult to break. We will revisit this in 6.1.

In the EHT analysis, GRMHD simulations are used ampliatively. This includes both model fitting to the M87* data, and identification of features detectable by proto-EHT arrays given priors from later observations. These exhibit theory-ladenness in their reliance on GRMHD simulations. If GRMHD simulations were inaccurate, this would introduce systematic bias to the inferences made about M87*.

However, this form of theory-ladenness seems to fall short of leading to problematic inferential circularity. In this context, inferential circularities might arise when source properties are used to test GR. However, there are strategies available for mitigating the epistemic risks of using GR in determining those properties. For example, there are alternative means of estimating some of the crucial parameters (such as stellar and gas dynamics); and some of the parameters might also be jointly constrained with other data (this happens for models of emission which rely on jet power as an additional input). Moreover, a variety of both GR-based as well as more exotic models of the central source can be considered and to some extent rejected—a topic we discuss in more detail in the next section.

The use of VLBI to image the "shadow" of the supermassive black hole candidate in M87 is less theory-laden than LVC observations. This is because the EHT imaging pipeline do *not* rely on GRMHD simulations, but rather on imaging algorithms that make few theoretical assumptions about the source. Moreover, in contrast to LVC, the EHT observations do not require substantial assumptions about GR (or alternatives to it) during data collection or imaging. However, much like the LVC, the EHT does rely on GRMHD simulations for parameter estimation. That part of the analysis can be performed over the same dataset, reducing reliance on GRMHDs.

Overall, the most severe instance of theory-ladenness we have discussed is the case of the LVC parameter estimation of compact binaries. This inference adds an additional layer of theory-ladenness on top of the theory-ladenness associated with matched filtering. What's more, this seems to be a case in which the theory-ladenness is inherent to the epistemic situation—the lack of independent empirical access to the target systems makes it more challenging to break the inferential circularity that arises in this context.

# 6 Mitigating the Risks of Theory-laden Observations

Zooming out from the details of the case studies, can anything more general be said about the ways of mitigating the epistemic risks of theory-ladenness? In other words, how can scientists increase the security of their inferences when these rely on heavily theory-laden methods?

In this section we discuss different strategies that scientists employ in contexts where there are concerns about the theory-ladenness of their methods. These strategies are quite general and may be deployed for a range of purposes. However, our goal in discussing them is just to show their usefulness in the context of theory-laden inference. In particular, possible biases introduced by theory-laden methods may be identified or ruled out through their applications. In our case studies from black hole astrophysics, this theory-ladenness arises from the ampliative use of simulation, so these strategies are of particular relevance when countering the epistemic risks of such ampliative approaches.

We identify five different types of strategies employed in the mitigation of inferential risks imposed by theory-ladenness: Independent Evidence; Weakening; Explicit; Generic Effect; and Parameterization. After explaining each of these strategies, we examine the strategies employed in our most problematic case study: LIGO-Virgo parameter estimation. Given the availability of mitigation strategies—even in this most challenging case—we think it is possible to take an optimistic stance toward theory-laden observations of black holes, even in cases where theory-ladenness leads to inferential circularity.

We then connect our analysis of these strategies, and of theory-laden experiments, to a parallel discussion of 'model independence' in the context of High Energy Physics (HEP). Some of the strategies we identify in this section have already received some philosophical attention in the HEP context. Tying these together with our discussion, a picture emerges in which the five strategies below can be understood as varying the model dependence (which we take to be a version of theory ladenness) across two dimensions: the degree to which the initial inference relies on a specific model (tight theoretical constraints), and the degree to which a range of alternative models are considered (broad exploration of models that relax theoretical constraints).

## 6.1 Five mitigation strategies

INDEPENDENT EVIDENCE STRATEGY: The first strategy involves using other means of accessing the system. For example, in both the M87* and Sgr A* cases, there are independent (from the EHT) methods for estimating the mass of the source. (For M87* these come from stellar dynamical measurements and gas dynamical measurements, cf. sections 8 and 9 of The Event Horizon Telescope Collaboration et al. (2019f); for Sgr A* the particularly important one utilizes observations of bright tracers such as star S2 (Ghez et al. 1998; Eckart and Genzel 1997).) Using this strategy, agreement across independent lines of evidence both bolsters the confidence in the outcome, and provides some grounds for thinking that the particular theoretical inputs are not essential for obtaining the result. Multi-messenger astrophysics is one version of this strategy, where a high level of independence across lines of evidence is secured by observing via different messengers (photons, neutrinos, gravitational waves, etc.).[15] However, some level of independence can also be secured by observing at different wavelengths, using different telescopes or arrays, varying the specific observational targets, etc.

WEAKENING STRATEGY: The second strategy is a methodological approach that Ritson and Staley (2021) call a weakening strategy. This is carried out by varying the theoretical assumptions used in a measurement in order to evaluate the systematic uncertainty. The reported conclusion is then "weakened", by implementing corresponding error bars. As a result, the measurement's "larger error bars make them compatible with a broader range of theoretical alternatives" (p. 155), reducing the reliance of the outcome on the particular choice of theoretical assumptions and mitigating potential for circularity. In particular, Ritson and Staley (2021) argue, using two ATLAS measurements, that in the context of High Energy Physics (HEP) uncertainty evaluation can disarm the viciousness of inferential circularity, by allowing a broader range of theoretical alternatives than the specific assumption used in the inference. In the context of the EHT a good example of this strategy is the resolution of the final M87* image. Different imaging pipelines are used to produce images, each with different resolution. These are then blurred to the most conservative one. (See Doboszewski and Elder (2024) for a philosophical analysis of this issue in the M87* imaging process.)

---

15. See Abelson (2022) and Elder (2024) for analyzes of multi-messenger astrophysics.

EXPLICIT STRATEGY: The third strategy involves explicit comparison to a particular well-determined theoretical alternative. This strategy involves a calculation of the specific observable for that alternative and comparing it with the data, to judge the viability of the alternative interpretation. In the EHT context, this is carried out as an (exploratory) comparison with both GR and non-GR alternatives, including some horizonless exotic compact objects (boson stars), naked singularities, and wormholes, as well as some black hole spacetimes of Einstein–Maxwell-dilaton-axion modified theory of gravity (see section 7.4 of (The Event Horizon Telescope Collaboration et al. 2019e) and (The Event Horizon Telescope Collaboration et al. 2022f)). Similarly, gravitational-wave templates have recently been constructed for some modified gravity theories (Ripley 2022); such templates might eventually be included in the modeled search pipelines (or in exploring different interpretations of detected gravitational waves, via parameter estimation), reducing the extent to which those are theory-laden with GR assumptions.

GENERIC EFFECT STRATEGY: The fourth strategy involves searching for generic effects associated with large classes of theoretical alternatives—in this case, non-black hole alternatives. For example, if the source under investigation was not a black hole—say, as characterized by an event horizon—but some kind of exotic object, it is likely to have a surface. Here, the exact nature of the exotic object is not specified in detail. Regardless of the physical origin of such a surface, this could potentially be detected by either LIGO-Virgo or the EHT. In the LIGO-Virgo case, gravitational wave signals could contain an additional component resulting from reflection of gravitational radiation off the surface of the other element of the compact binary. So far, the results of searches for these "echoes" are negative, but the search continues; see section 4.2 of Cardoso and Pani (2019) for a discussion of this effect. In the EHT case, the surface of this (non-black hole) exotic object should emit and reflect electromagnetic radiation. Signatures of this effect might be observable using VLBI methods; see Kleuver et al. (2023) for a further discussion.

PARAMETERIZATION STRATEGY: Finally, the fifth strategy involves performing a parameterization around a solution, with new parameters able to take on values that represent theoretical alternatives—including alternatives that are as-yet not well worked-out. One example of this is the parameterized post-Einsteinian ("ppE") framework of Yunes and Pretorius (2009) (see also philosophical discussion of this framework in Patton (2020)). This framework relaxes the assumption that general relativity provides the correct account of the generation and propagation of gravitational waves by constructing new gravitational waveform templates that include post-Einsteinian parameters (for example, sub-leading order corrections to the quadrupole formula).[16] This strategy can be seen as achieving, in part, what the Weakening strategy does in a different way (in that it results in showing compatibility with a broader range of theoretical alternatives) and also achieving some of the outcomes of the Generic Effect strategy (in that it captures classes of theories which do not have to be explicitly constructed). However, here the construction is more explicit and informative — at the price of introducing some theoretical assumptions (effective validity of GR templates), making it more dependent on those assumptions. To some extent, the Parametrization strategy avoids the problem of unconceived alternatives associated with the Explicit strategy, for which the question how to judge whether the data are compatible with alternative models in the situation where these models are not explicitly known is a hard problem.

Let's now consider an example: the case of LIGO-Virgo parameter estimation, which, we argued above, is the most severe instance of theory-ladenness across our case studies. Given that the parameter estimation is inherently model-dependent and that there are no independent means of empirical access to the binary black hole system, this might initially seem a hopeless case. The Independent Evidence strategy is inapplicable (for individual binary black hole mergers), because a single gravitational signal is all that we will ever have to work with.[17] The Weakening strategy is also unhelpful, because we expect (and *want*) parameter estimation to be sensitive to changes in the signal and the model—to reduce the dependence of the results on the particular theoretical assumptions would come at the cost of their being uninformative. However, the three remaining strategies—Explicit, Generic Effects, and Parameterization—are applicable.

For LIGO-Virgo parameter estimation, an ideal application of the explicit strategy would involve models mapping source parameters to signal morphology for a range of GR alternatives. Some first steps in this direction include e.g., the modified gravity templates of Ripley (2022) and the brief consideration of Chern-Simons-like corrections in Yunes and Pretorius (2009, 3).[18] An example of the Generic Effects strategy is the search for 'echoes' in the ringdown (post-merger) portion of the gravitational wave signal. Abedi, Dykaar, and Afshordi (2017) tried this strategy soon after the first detections were announced, and Cardoso and Pani (2019) offer a more recent report on it.

---

16. Note, following (Yunes and Pretorius 2009) section I.E, that even in a restricted setting, "the ppE construction is non-unique, and certainly more refined versions could be developed".

17. However, Elder (2024) discusses the role of independent evidence for multi-messenger observations of binary neutron star mergers.

Finally, a version of the Parameterization strategy based on Yunes and Pretorius (2009)'s ppE framework has been used extensively by the LIGO-Virgo collaboration (the 'parameterized deviations test') to determine the compatibility of detected signals with alternative models, without the need to have explicitly worked-out models on hand. This test looks for evidence that the gravitational wave signals favor non-GR parameter estimates over GR estimates. As Patton (2020, 146) puts it, "[Yunes and Pretorius] build a theory within the theory of GR, a parametrized set of models that allow for more rigorous testing of hypotheses about *deviations* from the theory's predictions and structure" (emphasis in original).

Overall, even this most challenging case of theory-ladenness and circularity looks more tractable when we consider these strategies. Even though LIGO-Virgo parameter estimation is inherently model-dependent, scientists can still mitigate the associated epistemic risks by probing for evidence that the data is compatible with alternative interpretations.

## 6.2 Model independence in astronomy and High Energy Physics

The five strategies discussed in section 6.1 involve varying degrees of model independence, in that they all explore alternatives to the theoretical assumptions made in the course of the original inference. The application of these strategies in black hole astrophysics has parallels with HEP, where theory-ladenness and circularity have similarly been topics of philosophical concern (Ritson and Staley 2021).

McCoy and Massimi (2018) have discussed the notion of model independence in the context of simplified models for physics beyond the Standard Model (BSM). A simplified model is an "extension of the SM that adds only a couple of new hypothetical BSM particles to the SM" (p. 104). In contrast to full BSM models, which contain complete descriptions of BSM physics, simplified models only have a small number of additional parameters. These models are unrealistic, in the sense that even if one of them was empirically adequate, it would still leave out a lot of BSM physics. But these models can nevertheless be useful. McCoy and Massimi mention three functions: simplified models can be used to interpret new data; evaluate and revise search strategies; and to compare data from different experiments. As a result, such models "allow experimentalists to learn something about the phenomena in a partially experiment- and data-independent way" (p. 116), which is of importance because data coming from a collider may be theory-laden, or, in McCoy and Massimi's own words, "may be dependent in various ways on the experimental apparatus and methodology" (p. 117).

---

18. However, the technical challenges of building these alternative models is part of the motivation for the use of the Parameterization strategy, which does not rely on specific knowledge of alternatives.

In a similar vein, King (2024) characterizes model independence as strategies aimed at reducing theory bias. In the HEP context, this usually refers to searches for a new physics based on deviations from the Standard Model, rather than on particular models of BSM physics. King sees model independence as a contrastive notion forming a spectrum, beginning with model-based searches, which attempt to test predictions derived from a particular model. Further on the spectrum are located moderately model independent searches (which King argues should instead be called model agnostic), relaxing the dependence on any particular single model, for example a simplified model of a supersymmetric quark; and, finally, approaches aiming to be as model independent as possible by significantly reducing modeling assumptions, involving e.g. fitting data to a machine learning model.

The terminology of model independence is also used in astrophysical discussions. For example, in a recent overview of possibilities of exploring fundamental physics with the LISA detector (Arun et al. 2022), a model dependent test is defined as one which compares the gravitational wave templates derived from a given theory of gravity (whether GR or some modified theory) to empirical data. In that paper, model independent tests are seen as coming in two types. The first is to test for inconsistencies between GR predictions and observations. For example, residuals tests (where the GR signal is subtracted from the data) are, in principle, sensitive to any discrepancies between the theory and the data. The second type attempts to perform model independent tests, and then map their outcomes to a particular modified gravity theory. The main test here relies on a parameterized formalism (inspired by the Parameterized post-Newtonian formalism), which contains "generic parameters capturing non-GR effects". Other model independent tests are also possible, including searches for gravitational wave polarization modes (scalar and vector) beyond those predicted by GR, or modified propagation of gravitational waves—in particular, differences in arrival times between gravitational and electromagnetic waves (these can be seen as an instance of our Generic Effect Strategy).

Similar to what King has argued for in the HEP case, the strategies for reducing the extent of theory-ladenness in black hole observations also form a spectrum of sorts. To us it seems that a line of investigation can be model independent in two distinct ways: with respect to their dependence on the particular set of general relativistic assumptions and with respect to the extent to which they depend on the specification of a particular theoretical alternative. (These two are logically distinct, making the metaphorical "spectrum" more complex.) In other words, model independence tracks how specific we are about determining both the starting point as well as the alternatives under consideration. Rather than being a property of the set of models, the "spectrum" of model independence manifests in different ways in which models are being used. How specific is the model we are considering? Is it a single model of the system (say, the Schwarzschild spacetime with a given mass), a set of models of the system (say, the full Kerr family); a specific way of modeling e.g. a compact binary, or a range of ways of modeling it? What assumptions, and how many of them, are they building in? Furthermore, are we considering a specific alternative that might model the data just as well as GR, or a range of different possibilities that could be alternatives? More freedom in specification of the alternative explanation makes the inference less theory laden and compatible with a wider range of alternatives; but it also tends to make the comparison less informative. Strategies which rely on heavily theory laden priors (for example, comparing specific models of GR with specific alternatives) can be more informative. In this sense, model independence involves trade-offs between informativeness (often associated with ampliative use of simulations relying on general relativistic assumptions) and the extent of theory ladenness.

# 7   Conclusions

We have extended Jacquart (2020)'s analysis of the ampliative role of simulations in astrophysics. Specifically, we have shown how some different epistemic challenges in black hole astrophysics lead to the use of simulations in an ampliative role, even when the dynamical timescales are short enough that astrophysicists can observe these systems as they change. We have also considered how using simulations to amplify observations relates to theory-ladenness and inferential circularity. This occurs to varying extent in the different examples we have examined, and has different consequences: some instances of theory-ladenness are relatively benign, while others (in particular, the LVC's parameter estimation) threaten to display inferential circularity. Given the ubiquity and the embeddedness of simulations in astrophysical observation, understanding the roles of simulations and their implications for issues of theory-ladenness across different contexts is crucial in providing a satisfactory account of the methodology and epistemology of both current and future observations in black hole astrophysics. However, in presence of the wide variety of strategies for reducing theory-ladenness and increasing model independence, one should be optimistic about the epistemic prospects of black hole astrophysical observations and their potential to discriminate between general relativistic black holes and both their known and modeled as well as underexplored, and perhaps even unknown and unconceived, alternatives.

## Acknowledgments

We are grateful to members of the Event Horizon Telescope collaboration and the LIGO-Virgo Collaboration, as well as to Nurida Boddenberg, Erik Curiel, Sam Fletcher, Sophia Haude, Martin King, Niels Martens, Helen Meskhidze, Jan Potters, Noah Stemeroff, and audiences in Pittsburgh, Munich, Bristol, London (Western Ontario), Brighton (Utah) for their feedback on previous versions of this paper.

We would also like to thank two anonymous reviewers for this journal for their useful comments and suggestions.

Both authors would like to gratefully acknowledge funding from the Lichtenberg Grant for Philosophy and History of Physics by the Volkswagen Foundation. This project/publication is funded in part by the Gordon and Betty Moore Foundation (Grant #8273.01). It was also made possible through the support of a grant from the John Templeton Foundation (Grant #62286). The opinions expressed in this publication are those of the author(s) and do not necessarily reflect the views of these Foundations.

Smith, George E., and Raghav Seth. 2020. *Brownian Motion and Molecular Reality: A Study in Theory-Mediated Measurement.* Oxford: Oxford University Press.

Tal, Eran. 2012. *The Epistemology of Measurement: A Model-Based Account.* http://search.proquest.com/docview/1346194511/.

———. 2016. "Making Time: A Study in the Epistemology of Measurement." 67 (1): 297–335.

The Event Horizon Telescope Collaboration et al. 2019a. "First M87 event horizon telescope results. I. The shadow of the supermassive black hole." *The Astrophysical Journal Letters* 875 (1): L1. https://doi.org/10.3847/2041-8213/ab0ec7.

———. 2019b. "First M87 event horizon telescope results. II. Array and instrumentation." *The Astrophysical Journal Letters* 875 (1): L2. https://doi.org/10.3847/2041-8213/ab0c96.

———. 2019c. "First M87 event horizon telescope results. III. Data processing and calibration." *The Astrophysical Journal Letters* 875 (1): L3. https://doi.org/10.3847/2041-8213/ab0c57.

———. 2019d. "First M87 event horizon telescope results. IV. Imaging the central supermassive black hole." *The Astrophysical Journal Letters* 875 (1): L4. https://doi.org/10.3847/2041-8213/ab0e85.

———. 2019e. "First M87 event horizon telescope results. V. Physical origin of the asymmetric ring." *The Astrophysical Journal Letters* 875 (1): L5. https://doi.org/10.3847/2041-8213/ab0f43.

———. 2019f. "First M87 Event Horizon Telescope Results. VI. The Shadow and Mass of the Central Black Hole." *The Astrophysical Journal Letters* 875, no. 1 (April): L6. https://doi.org/10.3847/2041-8213/ab1141.

———. 2021a. "First M87 Event Horizon Telescope Results. VII. Polarization of the Ring." *The Astrophysical Journal Letters* 910 (1): L12. https://doi.org/10.3847/2041-8213/abe71d.